# Mid-IR Metalens Based on MoS$_2$ Nanopillars

Muhammad Mahmudul Hasan
*Electrical and Computer Engineering Department*
*Florida International University*
Miami, USA
mhasa043@fiu.edu

Nezih Pala
*Electrical and Computer Engineering Department*
*Florida International University*
Miami, USA
npala@fiu.edu

*Abstract*—This paper proposes a metalens designed to work in 3.3 μm wavelength which resides in the mid-IR region. The metasurface was created using MoS$_2$ nanopillars taking advantage of high refractive index and low loss. It could be used to make compact optical Methane gas sensors.

*Keywords—component, formatting, style, styling, insert (key words)*

## I. Introduction

There is a growing interest about using the novel materials like chalcogenide to make optical instruments for the mid and long infrared region. Particularly, the peaks of the absorption profile of CH$_4$ and CO$_2$ gases resides at mid-IR region (~ 3.3 μm and~ 4.3 μm, respectively) [1]. Flat optics will help the gas sensors to be more compact and compatible. Metalens can be designed to work for targeted important wavelength spectrum like mid-wave and long-wave infrared spectrum. These lenses acquire $2\pi$ phase coverage depending on the lateral geometric structures while maintaining the same thickness/height. Metallic nanoantenna array based metalenses have the drawbacks of strong absorption loss and polarization dependency [2]. Dielectric resonators were investigated because of their low absorption loss [3], [4]. But typically to cover the full $2\pi$ phase shift these dielectric nanostructures are needed to be much thicker. III−V materials like GaP (n ~ 3.4) or AlGaAs (n ~ 3.8) could be a possibility to make the lens thinner at near-IR spectrum owing to their high refractive index and low absorption loss. However, due to the complexity to fabricate these material, Transition-metal dichalcogenides (TMDCs) and 2D materials could be a better alternative as they are independent of substrate and compatible to the modern fabrication technologies [5], [6].

Recently, it has been reported that high refractive index helps the optical path length (OPL) in 2D to increase which results in strong elastic light matter interaction [7]. TMDC materials especially MoS$_2$ could be potential candidate for exploring this area owing to its high refractive index ~4.2 [6], [8]. We proposed a design of metalens using MoS$_2$ nanopillars to explore the potential. The lens was simulated to work at 3.3 μm (mid-IR region) that can be used in compact optical CH4 gas sensor.

## II. Design Methodology

### A. Design Theory

The basic building blocks of metalens is the subwavelength lattice structure. These individual structures are responsible for specific required phase for that specific position of the metalens depending upon their geometrical dimension. Here, MoS$_2$ nanopillars are used as the subwavelength structures for the metalens design because of its high refractive index. Cylindrical shaped MoS$_2$ scatters can change the phase of wave depending on their height and diameter. The phase profile of a hyperbolic lens follows (1)

$$f(x, y) = \frac{2\pi}{\lambda} (f - \sqrt{x^2 + y^2 + f^2}) \quad (1)$$

Where, $\lambda$ represents the working wavelength for the lens, x and y are the position coordinates along the surface of the metalens and lastly $f$ is the focal length of the lens [4]. The nanopillars are comparable to shortened single-mode waveguide. Now for waveguide, the propagation phase of the wave when passing through the unit cell is as follows (2) [9].

$$f = \frac{2\pi}{\lambda} n_{eff} H \quad (2)$$

Here, $n_{eff}$ is the effective index which can be adjusted using the geometrical dimensions of the nanopillars and $H$ is the height of the unit cell structure. Apart from that, for a certain numerical aperture (NA), the periodicity of the unit cell should also follow the Nyquist sampling formula which will be,

$$P\ (periodicity) < \frac{\lambda}{2\ NA} \quad (3)$$

### B. Unit Cell

In our design, the height of the unit cell MoS$_2$ nanopillar is 7 μm and the periodicity of the unit cell is 900 nm which are in accordance with the formulas discussed above. Transparent SiO$_2$ is used as the substrate. The radius of the nanopillars is between 50 nm to 250 nm to cover the full $2\pi$ phase change. The structure of the unit cell of the metalens is showed in Fig. 1. The unit cell structure was simulated using the periodic boundary conditions in sideways in the FDTD simulation.

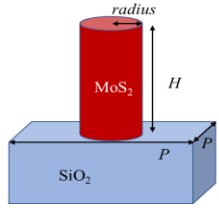

Fig. 1. Design structure of the unit cell of the metalens.

## III. SIMULATIONS AND RESULT ANALYSIS

FDTD numerical simulation was done using Ansys Lumerical. At first, we analyzed the design for the unit cell for the lens and its suitable dimensions. We have simulated the structure at wavelength 3300 nm (mid-wave IR). The structure is comprised of $MoS_2$ cylindrical pillar on top of glass substrate ($SiO_2$). The unit cell was simulated for different height lengths (5-9 µm) and for each height it was simulated for radius from 50 nm to 250 nm to evaluate the transmission and phase change profile. We noticed that the height required for the full phase coverage is larger than 2$\lambda$ of the concerned wavelength. Based on these results we have picked the optimum height (7 µm) for the $MoS_2$ nanopillar. After that, the radius of the pillar was varied from 50 nm to 250 nm keeping the height fixed at 7 µm to create the data set of phase change against the pillar radius which is depicted in Fig. 2. Data extracted from the unit cell analysis was then used to build the whole metalens.

After that, the focal length of the lens was determined to be 100 µm and the diameter of the lens was 60 µm. Considering the working wavelength and the numerical aperture of the lens the period of the unit cell was selected to be 900 nm preserving the Nyquist sampling criteria. The whole metalens surface was sampled by the period of 900 nm and required phase for each of those sampled space was calculated using (1). Based on the pillar radius vs. position in the metalens surface, the whole metalens structure was built and simulated for focal length of 100 µm. The far field analysis was done at the focal point which is shown in Fig. 3 displaying the field intensity distribution at focal point. The full-width half maximum (FWHM) of the lens is found to be ∼ 5 µm which is indicated in Fig. 3.

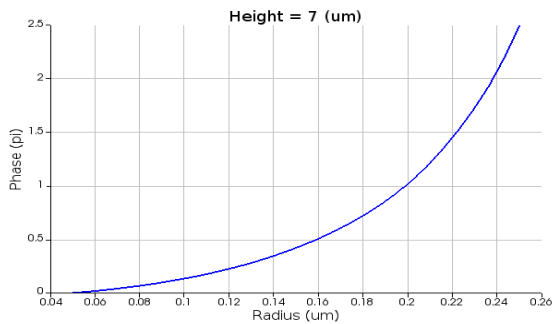

Fig. 2. $MoS_2$ nanopillar radius vs. phase change at height=7 µm.

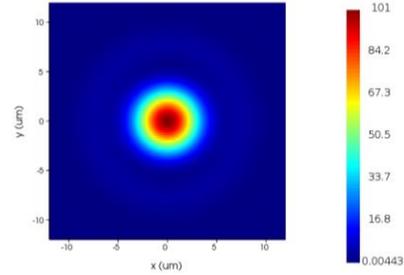

Fig. 3. Lens intensity at the focal point.

## IV. CONCLUSION

We successfully demonstrated a metalens design using nano cylindrical posts made of high refractive index material $MoS_2$. Most importantly, the lens was designed to work at 3.3-micron wavelength which is in the most important mid-wave IR region. The design focused on a simple approach to construct the lens while taking advantage of high refractive index of TMDC material. There is an issue about the required height of the pillars which might be a bit thick. But in future the lens can be made flatter applying some other techniques like using Huygens' meta-atoms or incomplete phase design approach. Apart from that the $MoS_2$ structures can be easily fabricated on a substrate material and most importantly this proposed metalens has the potential to contribute to many applications like optical sensing in the mid-wave IR range, imaging and energy harvesting etc.